\documentclass{aa} 
\usepackage{graphicx}
\usepackage{amsmath}
\usepackage{txfonts}
\usepackage{natbib}
\usepackage{hyperref}
\bibpunct{(}{)}{;}{a}{}{,}

\begin{document}

   \title{Meridional motions and Reynolds stress from SDO/AIA coronal bright points data\thanks{Table 1 is only available in electronic form at the CDS via anonymous ftp to \url{ftp://cdsarc.u-strasbg.fr} (\url{130.79.128.5}) or via \url{http://cdsweb.u-strasbg.fr/cgi-bin/qcat?J/A+A/}}}

   \author{D. Sudar
          \inst{1}
          \and
          S.~H. Saar
          \inst{2}
		  \and
          I. Skoki\'{c}
          \inst{3}
          \and
          I. Poljan\v{c}i\'{c} Beljan
          \inst{4}
          \and
          R. Braj\v{s}a
          \inst{1}
          }

   \institute{Hvar Observatory, Faculty of Geodesy,
              Ka\v{c}i\'{c}eva 26, University of Zagreb, 10000 Zagreb, Croatia
              \and
              Harvard-Smithsonian Center for Astrophysics, 60 Garden Street, Cambridge, MA 02138, USA
              \and
              Astronomical Institute of the Czech Academy of Sciences, Fri\v{c}ova
			  298, 251 65 Ond\v{r}ejov, Czech Republic
              \and
              Department of Physics, University of Rijeka, Radmile Matej\v{c}i\'{c} 2, 51000 Rijeka, Croatia 
             }

\offprints{D. Sudar, \email: davor.sudar@gmail.com}

   \date{Release \today}

\abstract 
{It is possible to detect and track coronal bright points (CBPs) in SDO/AIA images.
Combination of high resolution and high cadence provides a wealth of data that can be used
to determine velocity flows on the solar surface with very high accuracy.}
{We derived a very accurate solar rotation profile and investigated meridional flows, torsional oscillations
and horizontal Reynolds stress based on $\approx$6 months of SDO/AIA data.}
{We used a segmentation algorithm to detect CBPs in SDO/AIA images. We also used invariance
of the solar rotation profile with central meridian distance (CMD) to determine the height
of CBPs in 19.3 nm channel.}
{Best fit solar rotation profile is given by $\omega(b)=(14.4060\pm0.0051 + (-1.662\pm0.050)\sin^{2}b
+ (-2.742\pm0.081)\sin^{4}b)${\degr} day$^{-1}$. Height of CBPs in SDO/AIA 19.3 nm channel was found to
be $\approx$6500 km. Meridional motion is predominantly poleward for all latitudes, while
solar velocity residuals show signs of torsional oscillations. Horizontal Reynolds stress
was found to be small compared to similar works, but still showing transfer of angular momentum
towards the solar equator.}
{Most of the results are consistent with Doppler measurements rather than tracer measurements.
Fairly small calculated value of horizontal Reynolds stress might be due to the particular
phase of the solar cycle. Accuracy of the calculated rotation profile indicates
that it is possible to measure changes in the profile as the solar cycle evolves. 
Analysis of further SDO/AIA CBP data will also help in better understanding
of the temporal behaviour of the rotation velocity residuals, meridional motions and Reynolds stress.}

   \maketitle

\section{Introduction}
Studies of the solar rotation profile, torsional oscillations and meridional
velocities are based on either tracing specific features on or above the photosphere
or by using Doppler measurements.
The oldest known tracers for measuring rotation profile are sunspots which
have been used for a long time \citep{Newton1951, Howard1984, Balthasar1986,
Brajsa2002, Sudar2014}. The biggest advantage of sunspots is that they have
been observed for more than a century.
Coronal bright points (CBPs) have also been used very frequently by using
the data obtained by different satellites. For example, \citet{Brajsa2001, Brajsa2002b,
Brajsa2004, Vrsnak2003, Wohl2010} used SOHO/EIT data, \citet{Hara2009} analysed
Yohkoh/SXT measurements, while \citet{Kariyappa2008} used both
Yohkoh and Hinode data. Recently, \citet{Sudar2015} used SDO/AIA measurements
in 19.3 nm channel.

Doppler measurements showed similar results for rotation \citep{Howard1970, Ulrich1988, Snodgrass1990},
but analysis of meridional motions and torsional oscillation differ significantly
between tracer and Doppler measurements. Analyses of tracer data showed that meridional flow
is going out of the centre of activity \citep{Howard1986, Wohl2001},
while Doppler measurements usually show poleward meridional flow for all
latitudes \citep{Duvall1979, Hathaway1996}.
Of course, there have been studies that show the opposite. For example, \citet{Howard1991}
pointed out that solar plages show flow toward the centre of solar activity, unlike other
tracer measurements.
\citet{Perez1981} found the motion toward equator by analysing Doppler data in contrast
to other Doppler measurements.
\citet{Olemskoy2005} pointed out that for tracer measurements it is critical to 
take into account the distribution of the tracers in latitude in order not to
detect false flows.
Recently, \citet{Sudar2014} analysed sunspot group data from Greenwich Photoheliographic
Results and, by using the arguments from \citet{Olemskoy2005},
found that the meridional flow is toward the centre of solar activity.

\citet{Howard1980} reported that Sun is a torsional oscillator based on Doppler data.
This was later confirmed by \citet{Ulrich1988} again with Doppler measurements and \citet{Howe2000}
with helioseismic measurements. While many later papers found the torsional oscillation pattern
in such measurements, \citet{Sudar2014} were unable to detect anything like it in 150
years of sunspot group data.

Tracer data is very useful for analysis of horizontal Reynolds stress because
both velocity components can be measured separately \citep{Schroter1985}.
There have been a number of papers \citep{Ward1965, Schroter1976, Gilman1984,
Pulkkinen1998, Vrsnak2003, Sudar2014} which found the value of Reynolds stress
in agreement with transfer of the angular momentum toward the equator
which could explain the observed solar rotation profile.

\citet{Sudar2015} used SDO/AIA 19.3 nm channel to trace CBPs for two days.
Their results showed that the combination of high cadence/high resolution
satellite measurements can provide a wealth of data which could be used to analyse
variations of the solar rotation profile and all the associated phenomena
mentioned above. Analysis of meridional flow, torsional oscillations
and horizontal Reynolds stress with SDO/AIA CBP data is the main goal of this paper.

\section{Data and reduction methods}

In this work we used measurements from Atmospheric Imaging
Assembly (AIA) instrument which is on board Solar Dynamics
Observatory (SDO) satellite \citep{Lemen2012}.
We used a similar procedure as in our previous paper
\citep{Sudar2015} to obtain CBP positions. The segmentation algorithm is
a modification of similar algorithms described in
\citet{McIntosh2005} and \citet{Martens2012}.
In Table 1, available at CDS, we provide the following information.
Column 1 lists the Julian date of each observation, Column 2 contains
identification number of CBP, Column 3 and 4 give the $x$ and $y$ coordinates
for each CBP in pixels, respectively.

In order to obtain better accuracy than in \citet{Sudar2015},
where we used observational data from two days, in this work
we analysed more than 5 months of AIA/SDO observations with a 10
minute cadence from
2011 Jan. 1 -- 2011 May 19.
We removed data points near the limb ($>0.95R_{\odot}$) in order to
avoid potential problems with inaccuracies in position for those points.
Choosing only CBPs with 10 or more data points
to calculate velocities by linear fitting, we obtained 82341 velocity
measurements, which were converted from synodic to sidereal velocities
\citep{Skokic2014}.

In \citet{Sudar2015} we showed a change in position over time
of one CBP in Figure 2. Apart from the trend line, CBPs also exhibit
apparently random fluctuations around the fitted line. Such fluctuations
might be a result of displacements associated
with the evolution of CBP photospheric footpoints \citep{Karachik2014}.
Another cause might be due to image pixelation or changes in intensity
distribution of CBPs. Given that the spatial resolution of SDO/AIA is $\approx$0.6"/pixel
we can estimate that the error in position induced by CBP apparently
changing position by 1 pixel is about 0.04\degr in solar coordinates around
the equator. This is much smaller than observed by \citet{Sudar2015}.
The error in velocity between two subsequent images would be less than 1 m s$^{-1}$.
In our case, where we actually make a straight line fit through positions measured
in at least 10 images where the same CBP is detected, the error is even smaller than that.
Therefore, we assume that the observed fluctuations in CBP position are
most likely caused by the evolution of CBP photospheric footpoints described by \citet{Karachik2014}.
Since such fluctuations do not have some preferred direction on the solar surface,
we can assume that this effect averages out with a large number of data points.

With such a large number of data points obtained by an automatic
method, it is very likely that some velocities are wrong
due to misidentification in subsequent images or some similar problems.
It is quite common to filter out such outliers by selecting
a fixed range of acceptable rotational velocity \citep{Brajsa2002b, Vrsnak2003,
Sudar2014, Sudar2015}.
This approach neglects the fact that the solar rotation varies with latitude
and that such fixed cut-off does not have a uniform effect on all latitudes.
This in turn can affect the calculated rotation profile. Although, this effect
is probably negligible for the solar rotation, it might create problems
for derived quantities, such as rotation velocity residuals, or Reynolds stress.

\citet{Brajsa2001, Wohl2010} adopted a different, two steps, approach where they first
applied the fixed filter, calculated the solar rotation profile and then
eliminated all measurements which differed by more than 2{\degr} day$^{-1}$
from the calculated profile. Finally, the new profile was calculated
with a truncated dataset. This approach takes into account variation
of the rotation with latitude and is performing cut-off on rotation velocity
residuals.

We also developed a method which removes the outliers based on the
rotation velocity residuals trying to remove all arbitrariness
from the procedure.
The method we used is based on interquartile
range. First we calculate solar rotation profile from all data:
\begin{equation}
\label{Eq_rotProfile}
\omega(b)=A+B\sin^{2}b+C\sin^{4}b,
\end{equation}
where $b$ is the latitude and then we calculate rotation velocity
residuals.
Then we determine lower, $Q_{1}$, and upper quartile, $Q_{3}$, for the
rotation velocity residuals distribution.
We exclude all datapoints outside of the range:
\begin{equation}
[Q_{1} - k(Q_{3} - Q{1}), Q_{3}+k(Q_{3} - Q{1})],
\end{equation}
where we have chosen $k=3.5$ which removes so called hard outliers.
With the reduced dataset we calculate solar rotation profile again
and repeat the process iteratively until no datapoints are removed
by interquartile criterion.
In each iteration we removed the outliers by looking at meridional velocity
distribution with the same method.
The whole process is finished after only a few iterations.

Since CBPs are situated above the photosphere at unknown height,
we are actually measuring their apparent (projected) heliographic
coordinates \citep{Rosa1995, Rosa1998}.
To correct this problem we used the fact
that the solar rotation profile is invariant to central meridian distance (CMD).
We divided the solar disk into bins of 10{\degr} wide in CMD and calculated
the rotation profile for each bin getting a series of rotation profile
coefficients: $A_{i}$, $B_{i}$ and $C_{i}$ which can be compared
with the profile in the -5{\degr} to 5{\degr} CMD range defined by coefficients
$A$, $B$ and $C$. We can calculate
these coefficients for a number of different heights above photosphere
and request that the function:
\begin{equation}
\delta = \sum\limits_{i} \int_{0}^{\pi/2}\left(A_{i} - A + (B_{i}-B)\sin^{2}b
+ (C_{i}-C)\sin^{4}b\right)^{2}db,
\end{equation}
is minimal for some trial height, $h$. The integral is taken from the
equator to the pole so that the full profile is taken into account.
This integral can be evaluated since coefficients do not depend on the
latitude, $b$, so function $\delta$ becomes:
\begin{multline}
\delta = \sum\limits_{i} \biggl(\frac{\pi}{2}w_{A_{i}}(A_{i}-A)^2 + 
\frac{\pi}{4}\sqrt{w_{A_{i}}w_{B_{i}}}(A_{i}-A)(B_{i}-B)\\
+\frac{3\pi}{16}\left(w_{B_{i}}(B_{i}-B)^2 + \sqrt{w_{A_{i}}w_{C_{i}}}(A_{i}-A)(C_{i}-C)\right)\\
+\frac{5\pi}{32}\sqrt{w_{B_{i}}w_{C_{i}}}(B_{i}-B)(C_{i}-C) + \frac{35\pi}{256}w_{C_{i}}(C_{i}-C)^2\biggr),
\end{multline}
where we have introduced weights for coefficients $w_{A_{i}}$, $w_{B_{i}}$
and $w_{C_{i}}$ which are calculated from their errors obtained by fitting
the solar rotation profile in each CMD bin.
This height correction procedure was performed together with iterative
outlier removal process described above.

\begin{figure}
\resizebox{\hsize}{!}{\includegraphics{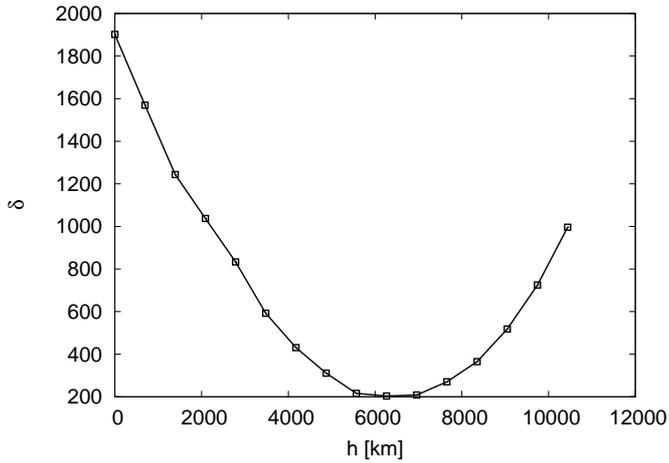}}
\caption{$\delta$ as a function of height, $h$, clearly
shows a minimum value at around 6500 km above the photosphere.}
\label{Fig_deltaOdH}
\end{figure}
By following the reasoning in \citet{Rosa1995} we can transform the apparent
coordinates into the deprojected ones by assuming that CBP are at some
height, $h$. This task is performed in polar coordinates obtained
from pixel coordinates \citep{Rosa1995}
so that both heliographic coordinates, CMD and latitude, are corrected for height.
We can then simply plot $\delta$ as
a function of $h$ and from the minimum detect the best fit height.
Such plot is given in Fig.~\ref{Fig_deltaOdH}. We can clearly see that
the minimum of function $\delta$ is located around 6500 km
giving us the average height of CBPs seen in SDO/AIA 19.3 nm channel.
By fitting the parabolic function to $\delta(h)$ we get the average height,
$h=6331\pm239$ km.

In the final run after all the filtering and with the best fit height we
had 80966 velocities in our dataset. For the analysis of rotation velocity residuals
and meridional motion we transformed the velocities to units of m s$^{-1}$.
We calculated meridional velocities on the southern hemisphere with
$v_{mer}=-\partial b/\partial t$ and assigned them symmetrical positive latitude.
This means that negative value of meridional velocity represents motion toward the
solar equator on both hemispheres.

\section{Results}
\subsection{Solar rotation profile and rotation velocity residuals}
In our previous paper \citep{Sudar2015} we estimated that with
5--6 months of SDO/AIA data we could obtain sufficient number of
velocity measurements that the accuracy
of the solar rotation
profile would be comparable with the most accurate tracer results so far.
Fitting the standard rotation profile (Eq.~\ref{Eq_rotProfile})
to 80966 measurements
we obtain for the coefficients: $A=14.4060\pm0.0051${\degr} day$^{-1}$,
$B=-1.662\pm0.050${\degr} day$^{-1}$ and $C=-2.742\pm0.081${\degr} day$^{-1}$.
It is important to point out that previous studies needed decades of measurements
to achieve this sort of accuracy.

As \citet{Snodgrass1985} explained, it is not straightforward to compare
result of the solar rotation profile from different sources when expressed as an expansion series
of $\sin^{2}b$ (Eq.~\ref{Eq_rotProfile}). To avoid crosstalk problem between
coefficients, it is better to express the result as Gegenbauer polynomials
which are orthogonal on the disk. Our solar rotation profile, expressed with
Gegenbauer polynomials is given by coefficients: $A_{G}=13.8386${\degr} day$^{-1}$,
$B_{G}=-0.698${\degr} day$^{-1}$ and $C_{G}=-0.131${\degr} day$^{-1}$.

Our result for the rotation profile is the most similar to the one by \citet{Hara2009}
who found the coefficients to be: $A=14.39${\degr} day$^{-1}$,
$B=-1.91${\degr} day$^{-1}$ and $C=-2.45${\degr} day$^{-1}$ or expressed
with Gegenbauer coefficients: $A_{G}=13.80${\degr} day$^{-1}$,
$B_{G}=-0.709${\degr} day$^{-1}$ and $C_{G}=-0.117${\degr} day$^{-1}$. \citet{Hara2009}
analysed X-ray bright points observed by the Yohkoh soft X-ray telescope
in the period 1994--1998. This time period starts close to the end of cycle 22
and ends soon after the beginning of cycle 23 \citep[see Table 1 in][]{Brajsa2009}.

Perhaps the similarity between our results and that of \citet{Hara2009}
are related to the low solar activity in both works.
For example, \citet{Brajsa2004} found a slightly higher value of the equatorial rotation
$A=14.454\pm0.027${\degr} day$^{-1}$ in the period from 1998-1999 which is closer to
the solar activity maximum. \citet{Wohl2010} found even faster equatorial rotation
$A=14.499\pm0.006${\degr} day$^{-1}$ with CBP data covering most of the cycle 23
around its activity maximum.
If such variations in the solar rotation profile are indeed due to the changing
activity of the sun, given the coefficient uncertainties we calculated above,  
we should be able to detect and track these changes
during the solar activity cycle with the expanded SDO/AIA CBP dataset.

\begin{figure}
\resizebox{\hsize}{!}{\includegraphics{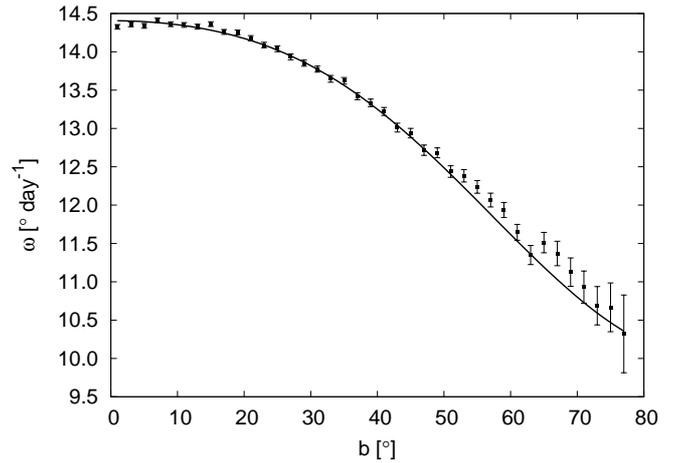}}
\caption{Average solar rotation profile in 2{\degr} bins of latitude, $b$, is shown
with filled black squares and error-bars.
The best fit solar rotation profile is shown with a solid black line.}
\label{Fig_rotProfile}
\end{figure}
In Fig.~\ref{Fig_rotProfile} we show the best fit rotation profile, $\omega(b)$, with a solid
black line. We also show average values
of $\omega$ in 2{\degr} bins of latitude, $b$, with with black squares and error bars. We see that
the bin averaged values are fairly well determined up to high latitudes (70\degr)
which is very promising for further CBP studies based on the SDO/AIA data.
Small size of the error-bars also illustrates how well the rotation profile
is determined.

\begin{figure}
\resizebox{\hsize}{!}{\includegraphics{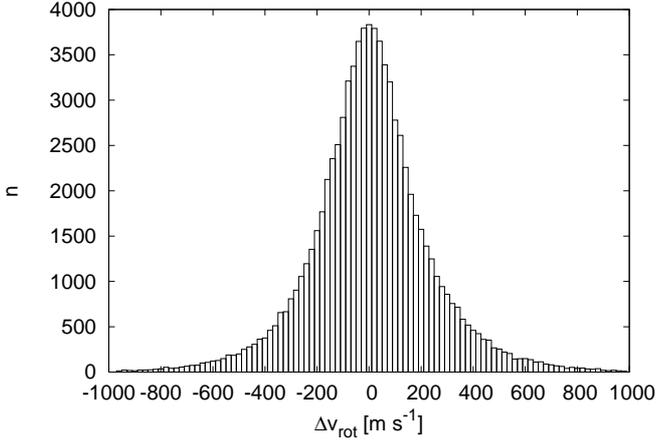}}
\caption{Distribution of rotation velocity residuals, $\Delta v_{rot}$, in bins 20 m s$^{-1}$
wide.}
\label{Fig_dvRotDistr}
\end{figure}
Rotation velocity residuals, $\Delta v_{rot}$, are calculated by subtracting actual rotation
velocity of each CBP from the mean profile given by the coefficient of the fit above.
The residuals are further transformed from units of {\degr} day$^{-1}$ to
m s$^{-1}$ where we took into account the latitude of each CBP.
In Fig.~\ref{Fig_dvRotDistr} we show a distribution of rotation velocity residuals, $\Delta v_{rot}$.
Since $\Delta v_{rot}$ was used to eliminate the outliers, it is important to check
if there are any unusual features in their distribution which would indicate that
something went wrong with our procedure. The distribution in Fig.~\ref{Fig_dvRotDistr}
looks fairly normal and well-behaved so we assume that the method we used is acceptable.

\begin{figure}
\resizebox{\hsize}{!}{\includegraphics{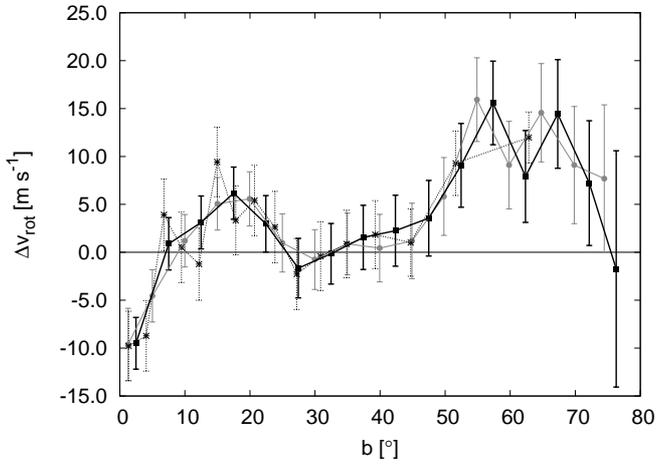}}
\caption{Rotation velocity residuals, $\Delta v_{rot}$,
in 5{\degr} bins in latitude, $b$, are shown with a thick black solid line and filled squares with error-bars.
Values calculated with bins of the same width but with the first bin in the range 0{\degr}--2.5{\degr} are shown
with a grey solid line and filled circles with error-bars. $\Delta v_{rot}$ averages in bins with constant number of
data points ($n$=5000) are shown with a dotted line and stars
with error-bars.}
\label{Fig_torsional}
\end{figure}
\citet{Tlatov2013} proposed an interesting idea that the torsional oscillation pattern,
associated with rotation velocity residuals, could, at least partially, be an
artefact of binning in latitude, $b$. The authors have been successful in simulating
the torsional oscillation pattern by assuming drifting of the tracers towards
the solar equator during the solar activity cycle. This is most notable
for sunspots with characteristic butterfly diagram, but could also be visible for CBPs.
Moreover, the authors suggest that this effect is present in Doppler and helioseismology
measurements.

In Fig.~\ref{Fig_torsional} we show the values of rotation velocity residuals,
$\Delta v_{rot}$, grouped into 5{\degr} bins of latitude, $b$, with a thick black solid line and black squares
with error-bars.
In order to address the problem of binning in latitude, we have also calculated
$\Delta v_{rot}$ averages in shifted bins, where the first bin is in the range from 0{\degr}
to 2.5{\degr} while all subsequent bins are 5{\degr} wide (shown with a grey solid line and filled circles
with error-bars in Fig.~\ref{Fig_torsional}).
In addition we also calculated $\Delta v_{rot}$ averages in bins with constant number of
data points ($n$=5000) and show the results in the same graph with a dotted line and stars
with error-bars.
We also calculated average latitude, $\bar{b}$, instead of using a middle value of $b$
for each bin. It should also alleviate the binning problem described
by \citet{Tlatov2013}, because the value of $\bar{b}$ is much more adequate in the case
of uneven distribution of data points in the bin.

From Fig.~\ref{Fig_torsional} we can conclude that all three binning techniques we used
show practically the same behaviour. We also want to point out that the values of
$\bar{b}$ do not differ significantly from the middle value of $b$ for each bin.
The difference is lower than 0.2{\degr} for all but one bin. This suggests that
the distribution of tracers in each bin is not far from uniform.
In Fig.~\ref{Fig_torsional} we see positive values of $\Delta v_{rot}$ averages in the
range between 10{\degr} and 20{\degr} which
might correspond to equatorward branch of torsional oscillation pattern.
Another branch might be visible above 50{\degr}, even though the errors are becoming
large in that region of latitudes. This type of result is more consistent with Doppler
measurements than with sunspot data where \citet{Sudar2014} found no stable pattern
which would resemble torsional oscillations.

\subsection{Meridional velocities}
In order not to detect false meridional flows due to uneven distribution
of tracers across different latitudes, it is necessary
to assign calculated velocities to the latitude of the
first measurements of position for each CBP \citep{Olemskoy2005, Sudar2014}.

\begin{figure}
\resizebox{\hsize}{!}{\includegraphics{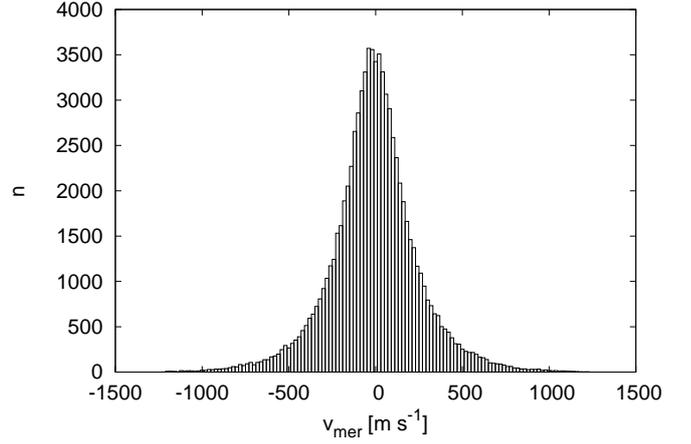}}
\caption{Distribution of meridional velocities, $v_{mer}$, in bins 20 m s$^{-1}$
wide.}
\label{Fig_meriDistr}
\end{figure}
As with $\Delta v_{rot}$, it is wise to take a look at the distribution of
meridional velocities, $v_{mer}$, (Fig.~\ref{Fig_meriDistr}) because they were also used in outlier
identification and elimination from the raw dataset.
Again, there are no unexpected features in the distribution which leads us to
believe that the method used was valid and correctly implemented.

\begin{figure}
\resizebox{\hsize}{!}{\includegraphics{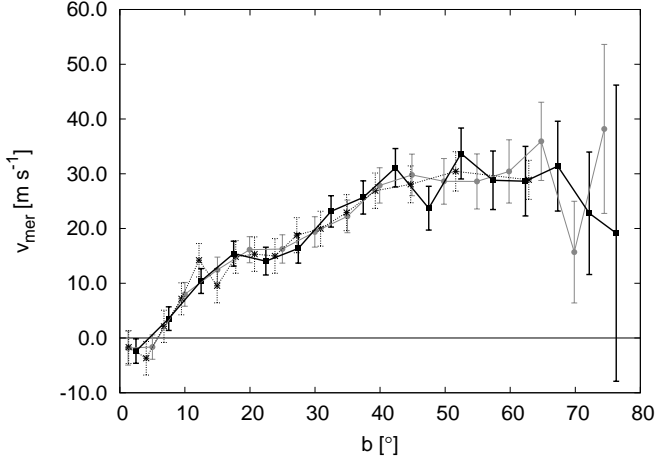}}
\caption{Average meridional velocity, $v_{mer}$ in 5{\degr} bins of latitude, $b$,
are shown with a thick black solid line and filled squares with error-bars.
Values calculated with bins of the same width but with the first bin in the range 0{\degr}--2.5{\degr} are shown
with a grey solid line and filled circles with error-bars. $v_{mer}$ averages in bins with constant number of
data points ($n$=5000) are shown with a dotted line and stars with error-bars.}
\label{Fig_meridional}
\end{figure}
In Fig.~\ref{Fig_meridional} we show average meridional velocity, $v_{mer}$, as a function
of latitude, $b$. As in the case with $\Delta v_{rot}$ we used three different binning
techniques and show the results with the same symbols as in Fig.~\ref{Fig_torsional}.
From the image we can see that the meridional velocity is almost always positive, meaning towards the poles,
for all latitudes. This result is in contrast to what was found with
sunspot groups \citep{Sudar2014} where the authors detected flow toward
the centre of solar activity for all latitudes and for all phases of the solar cycle.
Predominantly poleward flow was not found in other works dealing with tracers
\citep{Howard1986, Wohl2001, Vrsnak2003}.

On the other hand, poleward flow for all latitudes is detected by using the Doppler
method \citep{Duvall1979, Hathaway1996}. There is a small indication that for latitudes
near the equator the flow is equatorward. \citet{Snodgrass1996} already reported
that such feature is present in their analysis of Mt. Wilson magnetograms. Moreover,
they found that this low latitude behaviour is actually changing over the course
of the solar cycle. It will be very interesting to check if such behaviour is present
in the expanded SDO/AIA CBP dataset.

\subsection{Horizontal Reynolds stress}
Horizontal Reynolds stress is simply defined as a product
of rotation velocity residuals and meridional velocities
averaged over longitudes for a given latitude band:
\begin{equation}
q = <\Delta v_{rot}v_{mer}>.
\end{equation}
In our convention, if the value of $q$ is negative, it means that the angular momentum
is transported toward the solar equator.

\begin{figure}
\resizebox{\hsize}{!}{\includegraphics{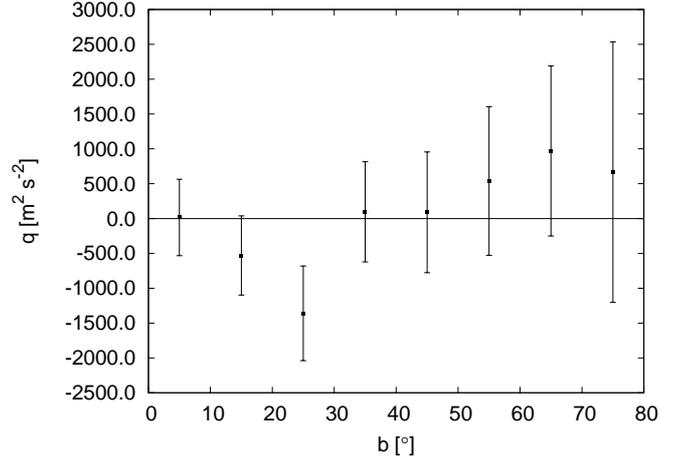}}
\caption{Horizontal Reynolds stress in bins of 10{\degr} in latitude, $b$, is
shown with filled black squares with error-bars}
\label{Fig_Reynolds}
\end{figure}
In Fig.~\ref{Fig_Reynolds} we show the value of the horizontal Reynolds
stress, $q$, as a function of latitude, $b$, with black squares. The bins are 10{\degr} wide
and the error-bars are also shown. We can see that the Reynolds stress
is zero for almost all latitudes, $b$. The only notable exception
is the bin at 25{$\degr$} where $q\approx$-1500 m$^{2}$ s$^{-2}$
and possibly the bin at 15{$\degr$}.
\citet{Sudar2014} found the minimum at the same latitude, but about
twice as deep. They also found negative values of $q$ for all
latitudes below 35{\degr}. Latitudes above 35{\degr} were out of
reach for sunspot groups measurements.

The minimum at 25{$\degr$} seems unexpected when we look at the behaviour
of average meridional velocity, $v_{mer}$, and rotation velocity residuals,
$\Delta v_{rot}$, at the same latitude. The two velocities show positive
average value, while horizontal Reynolds stress is negative.
If we take into account that the average of the product is not equal to the product
of averages, $<v_{mer}\Delta v_{rot}> \neq <v_{mer}><\Delta v_{rot}>$, we see
that the negative Reynolds stress we obtained actually means that the two velocities are
not independent, but correlated.

\section{Summary and Conclusion}
By using just under six months of SDO/AIA observations we have
calculated the solar rotation profile with accuracy comparable
to other tracer measurements which needed much longer time span
of observations.
Calculated solar rotation profile and comparison with other works
indicates that our result is connected with a low solar activity
in the observed phase of the current solar cycle. Further work
with more SDO/AIA data might provide some valuable insight about
the behaviour of the solar rotation during the solar cycle.

We found that CBPs observed by SDO/AIA 19.3 nm channel are located
at the average height of $\approx$6300 km above the solar photosphere.
This is slightly lower when compared to previous studies:
\citet{Simon1972} $\approx$11000 km, \citet{Brajsa2004} 8000 -- 12000 km
and \citet{Hara2009} $\approx$12000 km.
On the other hand, \citet{Karachik2006} suggest the value of 80000 km
which is the height at which \ion{Fe}{XIV} $\lambda$195 line forms \citep{Zhang2000}.
Rotation velocity residuals show indications of torsional oscillations
and further studies of the evolution of observed features might be very
helpful for comparison with other methods.

Meridional velocities are almost always towards the solar poles which
is what is often seen in helioseismology measurements
\citep{Zhao2004, Gonzales2008, Gonzales2010}. Observations of sunspot
groups, on the other hand, show a different meridional velocity pattern
\citep{Sudar2014}. However, \citet{Sudar2014} pointed out that
meridional velocity residuals in helioseismology measurements show
a striking similarity with sunspot groups observations.
The difference between CBPs and sunspot groups can be explained with similar
arguments as in \citet{Sudar2014} who suggested that sunspot observations
show motions related to active regions while the mostly poleward
flow is observed outside of those regions \citep[][Fig.~5]{Zhao2004}.
Our segmentation algorithm has difficulties detecting CBPs over bright
active regions, so CBPs results are more similar to time-distance heliosiesmology studies
than sunspot measurements.

Reynolds stress shows a minimum at around 25{\degr} in latitude similar to
results from \citet{Sudar2014}, but with lower magnitude ($q(25\degr)\approx$-1500
m$^{2}$ s$^{-2}$, compared to $\approx$-3000 m$^{2}$ s$^{-2}$ in \citet{Sudar2014}).
We are not sure if the reason for this result is the same as for the meridional velocities or
that this is some peculiarity of the phase of the solar cycle or even the
whole cycle 24.

Further work on the expanded SDO/AIA dataset and even possible application
of the segmentation algorithm to previous satellite measurements, such as SOHO/EIT,
will be very helpful for our understanding of the dynamics on and above
the photosphere. Such research can be considered complementary to helioseismology
measurements which probe the behaviour below the solar surface.

\begin{acknowledgements}This work has been supported in part by the Croatian Science Foundation
under the project 6212 "Solar and Stellar Variability".
It has also received funding from the SOLARNET project
(312495, 2013-2017) which is an Integrated Infrastructure Initiative (I3)
supported by FP7 Capacities Programme. SS was
supported by NASA Grant NNX09AB03G to the Smithsonian Astrophysical
Observatory and contract SP02H1701R from Lockheed-Martin to SAO.
\end{acknowledgements}

\bibliographystyle{aa}
\bibliography{sdo_q} 

\end{document}